# High-pressure evolution of $Fe_2O_3$ electronic structure revealed by x-ray absorption

Shibing Wang,[1,2,*] Wendy L. Mao,[2,3,4] Adam P. Sorini,[2] Cheng-Chien Chen,[2,5] Thomas P. Devereaux,[2,6] Yang Ding,[7] Yuming Xiao,[8] Paul Chow,[8] Nozomu Hiraoka,[9] Hirofumi Ishii,[9] Yong Q. Cai,[10] and Chi-Chang Kao[11,†]

[1]*Department of Applied Physics, Stanford University, Stanford, California 94305, USA*
[2]*SIMES, SLAC National Accelerator Laboratory, Menlo Park, California 94025, USA*
[3]*Photon Science, SLAC National Accelerator Laboratory, Menlo Park, California 94025, USA*
[4]*Department of Geological and Environmental Sciences, Stanford University, Stanford, California 94305, USA*
[5]*Department of Physics, Stanford University, Stanford, California 94305, USA*
[6]*Geballe Laboratory for Advanced Materials, Stanford University, Stanford, California 94305, USA*
[7]*HPSynC, Carnegie Institution of Washington, Washington, DC 20015, USA*
[8]*HPCAT, Carnegie Institution of Washington, Washington, DC 20015, USA*
[9]*National Synchrotron Radiation Research Center, Hsinchu 30076, Taiwan*
[10]*NSLS-II, Brookhaven National Laboratory, Upton, New York 11973, USA*
[11]*SSRL, SLAC National Accelerator Laboratory, Menlo Park, California 94025, USA*

We report the high-pressure measurement of the Fe $K$ edge in hematite ($Fe_2O_3$) by x-ray absorption spectroscopy in partial fluorescence yield geometry. The pressure-induced evolution of the electronic structure as $Fe_2O_3$ transforms from a high-spin insulator to a low-spin metal is reflected in the x-ray absorption pre-edge. The crystal-field splitting energy was found to increase monotonically with pressure up to 48 GPa, above which a series of phase transitions occur. Atomic multiplet, cluster diagonalization, and density-functional calculations were performed to simulate the pre-edge absorption spectra, showing good qualitative agreement with the measurements. The mechanism for the pressure-induced electronic phase transitions of $Fe_2O_3$ is discussed and it is shown that ligand hybridization significantly reduces the critical high-spin/low-spin transition pressure.

## I. INTRODUCTION

An archetypal $3d$ transition-metal oxide and important geological compound, $\alpha$-$Fe_2O_3$ (hematite) undergoes a series of structural and electronic transitions at high pressure. At ambient conditions, $Fe_2O_3$ is an antiferromagnetic insulator and adopts the corundum structure. This structure is maintained until approximately 50 GPa whereupon it transforms to a $Rh_2O_3$(II)-type structure,[1] accompanied by a 10% drop in volume. The structural transition is associated with changes in magnetic and electronic structures. X-ray $K_\beta$ emission at ambient pressure and 72 GPa show that the magnetic moment drops from high spin (HS) to low spin (LS) at high pressure.[2] Conductivity measurements indicate that an insulator to metal transition occurs between 40 and 60 GPa.[3] Mössbauer spectroscopy up to 82 GPa (Ref. 3) and synchrotron Mössbauer spectroscopy at 70 GPa (Ref. 4) imply the collapse of the magnetic moments and a nonmagnetic nature of the HP phase.

The nature of these transitions has been a popular research topic over the past decade. Based on their structural study of the $Rh_2O_3$-II phase, Rozenberg *et al.*[1] have suggested that the charge-transfer gap closure is responsible for metallization and concurrent spin moment transition. Combined local-density approximation and dynamical mean-field theory calculations by Kuneš *et al.*[5] have implied that the reduction in the Mott gap with pressure drives the volume collapse and structure change. This idea appears to be at odds with experimental observations of a metastable state in which the HS and high-pressure structure occur simultaneously.[6] Thus, despite many studies of the transitions in $Fe_2O_3$, the nature of the evolution of the electronic structure with pressure remains unresolved. In this paper, we implemented experimental method and theoretical approaches bringing valuable information to the problem.

A number of spectroscopic techniques have been applied to investigate the electronic configuration of $3d$ transition-metal compounds. Photoemission and x-ray $L$-edge absorption provide useful information on the $3d$ levels of transition metals but unfortunately, these probes cannot penetrate the high pressure cells. X-ray absorption spectroscopy (XAS) at the $K$ edge of $3d$ transition elements, however, operates in the hard x-ray regime, allowing the study of the electronic structure at high pressure.

The pre-edge region of the $K$-edge absorption spectrum can be used to investigate $3d$ electrons of transition-metal compounds. In Fe-bearing compounds, the pre-edge spectra contain information about the oxidation state and local coordination.[7] However, limited by the $1s$ core-hole lifetime broadening, the energy resolution using a transmission geometry is not sufficient to resolve the detailed structure of the pre-edge region. Therefore we use the partial fluorescence yield method for measuring absorption. Instead of collecting the transmitted x-ray, the $K_{\alpha1}$ emission line is measured. This method thus has a $2p$ core-hole lifetime broadening of about 0.3 eV, resulting in much higher energy resolution.

Here we present the first high-pressure XAS measurement in partial fluorescence yield on $Fe_2O_3$ up to 64 GPa. The improved resolution of the resulting spectra shows the evolution of the $Fe^{3+}$ $3d$ electronic structure as the material un-dergoes its complex pressure-induced transitions. Previously, Caliebe *et al.* applied this technique to $Fe_2O_3$, and assigned the double-peak structure of the pre-edge to the $t_{2g}$ and $e_g$ components of the $3d$ band[8] as suggested previously.[9] Similar methods have been used to study orbital hybridization and spin polarization of $Fe_2O_3$ (Ref. 10) and pre-edges of other Fe-containing compounds.[11]

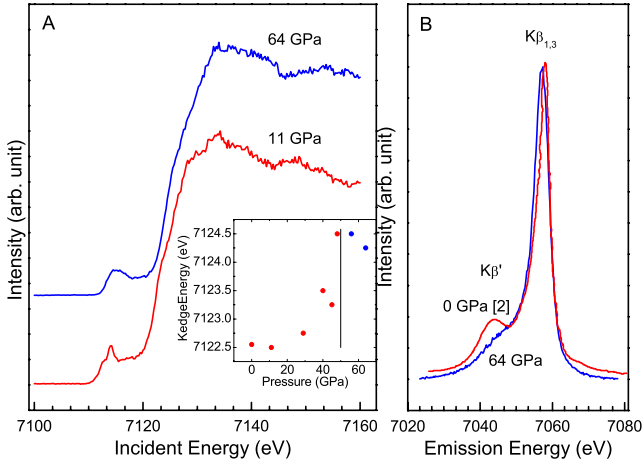

FIG. 1. (Color online) (A) X-ray $K$-edge absorption spectra of $Fe_2O_3$ in partial fluorescence yield geometry at 11 and 64 GPa; Inset: Fe $K$-edge position at different pressures. The edge is determined by the maximum of the first derivative of the absorption spectra. (B) X-ray $K_\beta$ emission spectra of $Fe_2O_3$ at 64 and 0 GPa from Ref. 2, showing the reduction in the spin moment. Red: high-spin state and blue: low-spin state.

## II. EXPERIMENT

$Fe_2O_3$ powder was loaded in a hydrostatic pressure transmitting medium (He or Ne) in an x-ray transparent Be gasket. Ruby fluorescence was used for pressure calibration. High-pressure XAS spectra of $Fe_2O_3$ were collected at two-third generation synchrotron facilities. In both setups, monochromatic x-rays focused by Kirkpatrick-Baez mirrors were directed through a panoramic diamond-anvil cell, and the analyzer was fixed at 90° from the incident beam.

In the SPring-8 XAS experiment conducted at BL12XU, we scanned the incident x-ray energy from 7110 to 7145 eV with a step size of 0.1 eV and over the smaller range of 7112–7115 eV at 0.05 eV step size. In the APS setup at HPCAT 16-IDD, the entire edge was scanned from 7100 to 7160 eV with a step size of 0.25 eV. The pre-edge was scanned from 7108 to 7118 eV (7109 to 7119 eV for 56 and 64 GPa) with a step size of 0.2 eV. For both measurements, the partial fluorescence yield was collected with the analyzers set at the Fe $K_{\alpha 1}$ energy (6405.6 eV).

Figure 1(A) shows the representative Fe $K$-edge XAS spectra for $Fe_2O_3$. The partial fluorescence yield geometry allows us to resolve the pre-edge features. At the highest pressure in our study, we collected the $K_\beta$ emission spectrum of the sample shown in Fig. 1(B). Compared with the 0 GPa spectrum of Badro *et al.*, there is a dramatic reduction in the $K'_\beta$ satellite peak intensity in the 64 GPa spectrum, indicating a LS ground state.[2,12]

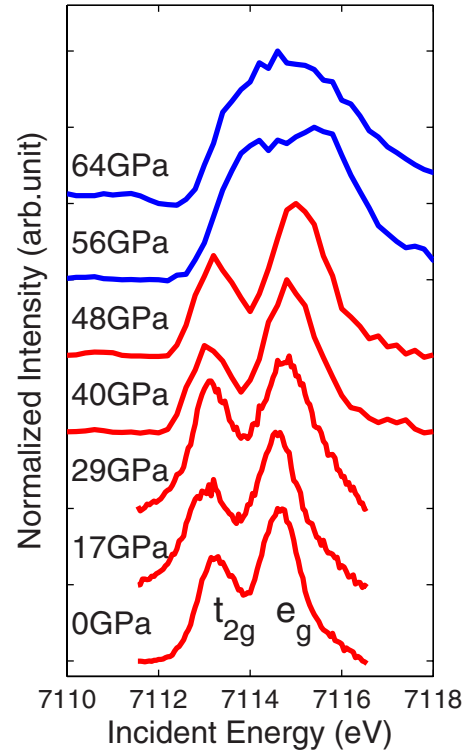

FIG. 2. (Color online) X-ray $K$-edge pre-edge of $Fe_2O_3$ at 0, 17, 29, 40, 48, 56, and 64 GPa. The bottom three spectra are from SPring-8 using high-resolution monochromator and the top four spectra are from APS using diamond monochromator.

As shown in Fig. 1(A) inset, it is also observed that the $K$-edge blueshifts with pressure until the phase-transition region and remain approximately constant thereafter. This shift of $K$ edge with pressure is also observed in other $3d$ transition-metal oxides,[13] a result of the increase in electron density upon compression.

Figure 2 shows the Fe $K$-edge pre-edge spectra of the sample from ambient pressure to 64 GPa. The tail of the main absorption edge was subtracted for each spectrum by removing the $K$-edge absorption spectrum of Fe in the Be gasket. The pre-edge features at ambient pressure are associated with excitations to $t_{2g}$ and $e_g$ orbitals, split by the octahedral crystal field. Our ambient pressure data can be fit with a crystal-field splitting energy (CFSE) of 1.4 eV, consistent with previous observation.[8,9] The two-peak feature in the pre-edge persists until 48 GPa, just before the phase transitions occur. By fitting the pre-edge spectra we estimate a monotonic increase in the CFSE to 1.85 eV at 48 GPa, as shown in Table I. This increase is expected as the $FeO_6$ octahedra shrink with pressure, and the shorter metal-ligand distance elevates the $e_g$ level relative to the $t_{2g}$ level.

The pre-edge spectra above the phase transitions (i.e., above 48 GPa) are more complicated to interpret. The full width at half maximum of the pre-edge features significantly broadens and a simple assignment in terms of single particle $t_{2g}$ and $e_g$ transitions is inconsistent; at such pressures, $Fe_2O_3$ is in the LS state in which $e_g$ should be empty and five of the six $t_{2g}$ states occupied. Such a single-particle configuration should lead to relatively small (large) $t_{2g}(e_g)$ amplitudes, un-like the features observed in the pre-edge spectra at 56 and 64 GPa.

TABLE I. Crystal-field splitting energy (CFSE) of Fe$_2$O$_3$ as a function of pressure.

| Pressure (GPa) | 0 | 6 | 17 | 29 | 40 | 48 |
|---|---|---|---|---|---|---|
| CFSE (eV) | 1.41 | 1.44 | 1.59 | 1.73 | 1.82 | 1.85 |

## III. THEORETICAL INTERPRETATION

To understand the pressure dependence of the XAS, we first used crystal-field atomic multiplet theory to calculate the electronic structure. The relevant parameters are the atomic $t_{2g}$-$e_g$ energy-level spacing $10Dq$ (Ref. 14) and the "Racah parameters" $B$ and $C$ associated with $d$-$d$ interactions.[15] We fix Racah $B=0.075$ eV and $C=0.346$ eV appropriate for solid-state Fe$^{3+}$ systems,[16] and perform calculations for a range of $10Dq$. The lowest two eigenenergies for the $(1s)^2(3d)^5$ configuration are shown in Fig. 3(a) from which a HS-LS transition is evident near $10Dq=2.2$ eV. For low pressure (low $Dq$) the ground state has $^6A_1$ character (HS) and crosses over at high pressure to a state of $^2T_2$ character (LS).[17,18]

While the critical value of $10Dq$ determined by the atomic multiplet calculation is larger than that suggested by the experimental $t_{2g}$-$e_g$ peak splitting in Fig. 2, it is well known that the critical $10Dq$ for the HS-LS transition is reduced by the Fe-O charge-transfer processes. We perform calculations on a FeO$_6$ octahedral cluster that explicitly includes multiplets, ligand hybridization and charge-transfer via the Slater-Koster matrix elements,[19,20] Racah parameter $A$, and charge-transfer gap energy $\Delta$. At ambient pressure, the values of the parameters are (in units of electron volt): $V_{pd\sigma}=-1.13$, $V_{pd\pi}=0.65$, $V_{pp\sigma}=0.56$, and $V_{pp\pi}=-0.16$, $A=5.0$, $10Dq=0.96$, and $\Delta=2.7$.[19] We have used the smaller value of $V_{pd\sigma}$ from Ref. 19. The lowering of the critical $10Dq$ is illustrated in Fig. 3(a), which shows the energies of the HS and LS states calculated in the FeO$_6$ cluster compared to atomic multiplet theory as a function of $10Dq$. The HS to LS transition occurs at smaller $10Dq$ since the hybridization most strongly couples the $d^5$ LS state with the $d^6\underline{L}$ LS state, lying lower in energy than the $d^6\underline{L}$ HS state.

These parameters yield the ambient pressure spectra shown in Fig. 3(c), which is in good agreement with experiment (cf. Fig. 2 and Table I). The two spectral peaks separated by ~1.4 eV correspond to excitations into the $t_{2g}$ and $e_g$ orbitals, respectively, and indicate a HS ground state, with the observed CFSE coming from $10Dq$ plus a 0.45 eV covalent contribution. Thus while the critical $10Dq$ is reduced by the Fe-O charge-transfer processes, the ligand field splitting due to covalency pushes up the spectral $t_{2g}$-$e_g$ peak separation of the XAS spectra.[8]

With parameters set to reproduce ambient spectra, we consider the pressure evolution of the HS-LS transition and the XAS spectra. As the pressure increases, both $10Dq$ and the hopping integrals increase, respectively having $\sim d^{-5}$ and $\sim d^{-4}$ Fe-O bond-length dependence.[14,19] The combined effect of pressure-dependent hopping and $10Dq$ is explained in the phase diagram of Fig. 3(b). We consider several variations in $V_{pd\sigma}$ with $d$ as shown in Fig. 3(b), which all indicate that the critical pressure occurs between 52 and 55 GPa. Although variation in the exponent of $V_{pd\sigma}$ induces a variation on the order of 5% in the predicted critical pressure it is striking to observe that the experimentally observed limits on the critical pressure are in general agreement with theoretical predictions.

Figures 3(c)–3(e) show the calculated pre-edge XAS spectra from the FeO$_6$ cluster at various pressures. The spectrum at 48 GPa [Fig. 3(d)] shows a clear two peak structure in the HS state, with a $t_{2g}$-$e_g$ peak separation of ~1.6 eV. The calculated CFSE is ~15% smaller in energy than experiment, which may be in part due to structural deviations from

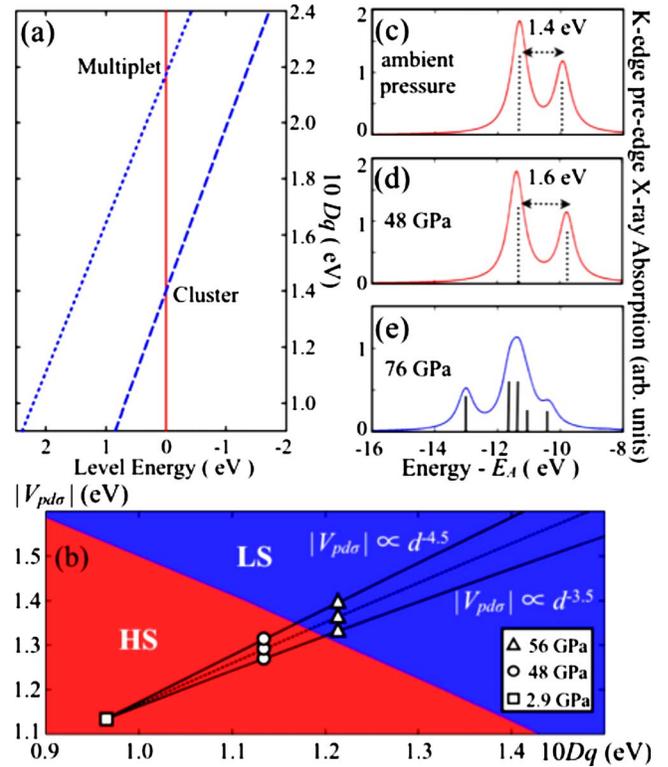

FIG. 3. (Color online) (a) Energy of LS state for the single atom multiplet calculation (dotted line) compared with the FeO$_6$ cluster diagonalization (dashed line) relative to the HS state (solid line). (b) HS-LS phase diagram for Fe$_2$O$_3$. The dotted line shows the probable trajectory of $(10Dq, V_{pd\sigma})$ with increasing pressure (see text). [(c)–(e)] $K$-edge pre-edge XAS spectra from the FeO$_6$ cluster calculation at various pressures; $E_A$ is the Fe $K$-edge absorption energy. (c) At ambient pressure, the spectrum shows distinct $t_{2g}$-$e_g$ absorption peaks separated by 1.4 eV, indicating a high-spin ground state. (d) At 48 GPa, the peak separation is 1.6 eV, and the ground state still resides in the high-spin sector. (e) At 76 GPa, the spectrum shows broad, multiple peaks, indicating a low-spin ground state. All the spectra were broadened with a 0.3 eV Lorentzian.

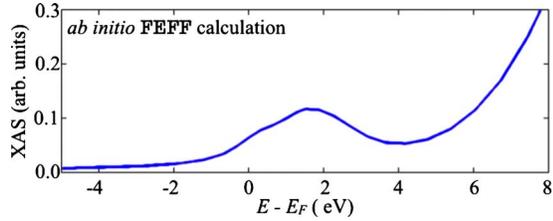

FIG. 4. (Color online) *Ab initio* calculations with the FEFF software of the pre-edge region of the Fe *K*-edge XAS spectra for the high-pressure metallic $Rh_2O_3(II)$-type structure of $Fe_2O_3$. $E_F$ is the Fermi level.

octahedral symmetry giving inequivalent Fe-O bonds not included in the cluster calculation,[1] as well as the overall uncertainty in cluster parameters. Figure 3(e) shows the calculated XAS spectra at 76 GPa. The high-pressure spectra have multiple-peak features indicating a LS ground state; this qualitative change in character of the ground state is reflected as a qualitative change in the calculated spectra. It is the simple transformation properties ($A_1$) of the HS state that allow the XAS to be interpreted in terms of single-particle $t_{2g}$ and $e_g$ levels; the final state, with one additional $d$ electron, transforms as $A_1 \otimes (T_2 \oplus E) = T_2 \oplus E$ mimicking the single-particle $t_{2g}$ and $e_g$ levels. On the other hand, addition of a $d$ electron to the LS state yields $T_2 \otimes (T_2 \oplus E) = A_1 \oplus E \oplus T_1 \oplus T_2 \oplus T_1 \oplus T_2$ resulting in more peaks than would be expected based on a single-particle interpretation.

While an insulator-metal transition is not necessarily concomitant with a change in the local-spin configuration (and vice versa), a low-spin metallic state is always expected at a high enough pressure. In this regime, we use the all-electron FEFF code[21–23] to calculate the high-pressure Fe *K*-edge XAS for a large cluster of 152 atoms in the high-pressure structure. Figure 4 shows the calculated pre-edge XAS, having broad pre-edge features in qualitative agreement with the experiment at and above 56 GPa.

We last turn to the electronic phase-transition mechanism. Badro *et al.* have shown the coexistence of HS and $Rh_2O_3$-II structure indicating that the electronic transition cannot drive the structural transition. Kuneš *et al.* divided the electronic transition into a Mott gap closing and a HS-LS gap closing, and estimated the respective regimes of stability via a local "density-based interaction." Here we have indicated the importance of atomic multiplets and ligand hybridization. Our results indicate the location of the HS-LS transition can be well described within the charge-transfer multiplet-hybridization cluster approach and reasonable choices for the pressure dependence of the cluster parameters. The reduction in the critical pressure for the HS-LS transition in comparison with atomic multiplet theory due to ligand hybridization is seen to be significant. These results lead to the prediction that the critical pressure occurs between 52 and 55 GPa, at values of $10Dq$ much smaller than would be expected from atomic multiplet theory based on the experimental spectra. While our cluster calculation cannot address in detail the closing of a bulk Mott gap, the observed reduction in the HS-LS transition pressure leads us to suggest that the physics of a local HS-LS transition should be strongly reconsidered as the key ingredient giving the evolution of spectral features observed in the pre-edge XAS spectra with pressure.

In summary, we measured x-ray absorption spectra of $Fe_2O_3$ up to 64 GPa, and experimentally resolved the crystal-field splitting and its pressure dependence through the metal-insulator transition. The CFSE increases from 1.41 eV at ambient conditions to 1.85 eV at 48 GPa. The pre-edge features change drastically at higher pressures corresponding to the range where a number of electronic and structural transitions have been reported. We constructed the phase diagram for $Fe_2O_3$ which shows that the changes in multiplet structure and hybridization are important for a quantitative estimate of the critical pressure. Based on considerations of local cluster physics, excellent agreement between the observed pressure dependence of the experimental and calculated spectra were obtained.

## ACKNOWLEDGMENTS

The authors thank W. Harrison, S. Johnston, E. Kaneshita, and B. Moritz for helpful discussions, and thank H.-k. Mao and J. Shu on experiments. S.W. and W.L.M. are supported by the NSF-Geophysics under Grant No. EAR-0738873 and Department of Energy through DOE-NNSA(CDAC) under Grant No. DE-AC02-76SF00515. A.P.S., C.C.C., and T.P.D. are supported by the U.S. Department of Energy under Contracts No. DE-AC02-76SF00515 and No. DE-FG02-08ER46540 (CMSN). Y.D. is supported by EFree funded by DOE (Grant No. DE-SC0001057). The experiment at SPring-8 was performed under the approval of JASRI (Grant No. 2007A4264) and NSRRC (Grant No. 2006-3-112-3). Portions of this work were performed at HPCAT (Sector 16), Advanced Photon Source (APS), Argonne National Laboratory. HPCAT is supported by DOE-BES, DOE-NNSA, NSF, and the W.M. Keck Foundation. APS is supported by DOE-BES under Contract No. DE-AC02-06CH11357.

---

*Corresponding author; shibingw@stanford.edu

†Also at NSLS, Brookhaven National Laboratory, Upton, New York 11973, USA.

[1] G. Kh. Rozenberg, L. S. Dubrovinsky, M. P. Pasternak, O. Naaman, T. Le Bihan, and R. Ahuja, Phys. Rev. B **65**, 064112 (2002).

[2] J. Badro, V. V. Struzhkin, J. Shu, R. J. Hemley, H.-k. Mao, C.-c. Kao, J.-P. Rueff, and G. Shen, Phys. Rev. Lett. **83**, 4101 (1999).

[3] M. P. Pasternak, G. Kh. Rozenberg, G. Yu. Machavariani, O. Naaman, R. D. Taylor, and R. Jeanloz, Phys. Rev. Lett. **82**, 4663 (1999).

[4] S.-H. Shim, A. Bengtsonb, D. Morganb, W. Sturhahnc, K. Catallia, J. Zhaoc, M. Lerchec, and V. Prakapenkae, Proc. Natl. Acad. Sci. U.S.A. **106**, 5508 (2009).

[5] J. Kuneš, Dm. M. Korotin, M. A. Korotin, V. I. Anisimov, and P. Werner, Phys. Rev. Lett. **102**, 146402 (2009).

[6] J. Badro, G. Fiquet, V. V. Struzhkin, M. Somayazulu, H.-k. Mao, G. Shen, and T. Le Bihan, Phys. Rev. Lett. **89**, 205504 (2002).

[7] M. Wilke, F. Farges, P.-E. Petit, G. E. Brown, Jr., and F. Martin, Am. Mineral. **86**, 714 (2001).

[8] W. A. Caliebe, C.-C. Kao, J. B. Hastings, M. Taguchi, A. Kotani, T. Uozumi, and F. M. F. de Groot, Phys. Rev. B **58**, 13452 (1998).

[9] G. Dräger, R. Frahm, G. Materlik, and O. Brümmer, Phys. Status Solidi B **146**, 287 (1988).

[10] P. Glatzel, A. Mirone, S. G. Eeckhout, M. Sikora, and G. Giuli, Phys. Rev. B **77**, 115133 (2008).

[11] J.-P. Rueff, L. Journel, P.-E. Petit, and F. Farges, Phys. Rev. B **69**, 235107 (2004).

[12] J.-P. Rueff and A. Shukla, Rev. Mod. Phys. **82**, 847 (2010).

[13] A. Y. Ramos, H. C. N. Tolentino, N. M. Souza-Neto, J.-P. Itié, L. Morales, and A. Caneiro, Phys. Rev. B **75**, 052103 (2007).

[14] J. H. Van Vleck, J. Chem. Phys. **7**, 72 (1939).

[15] G. Racah, Phys. Rev. **62**, 438 (1942).

[16] D. M. Sherman and T. David Waite, Am. Mineral. **70**, 1262 (1985); S. Brice-Profeta, M.-A. Arrio, E. Tronc, N. Menguy, I. Letard, C. Cartier dit Moulin, M. Nogués, C. Chanéac, J.-P. Jolivet, and Ph. Sainctavit, J. Magn. Magn. Mater. **288**, 354 (2005).

[17] Y. Tanabe and S. Sugano, J. Phys. Soc. Jpn. **9**, 766 (1954).

[18] J. Zaanen, G. A. Sawatzky, and J. W. Allen, Phys. Rev. Lett. **55**, 418 (1985).

[19] W. A. Harrison, *Elementary Electronic Structure* (World Scientific, Singapore, 2004).

[20] The experimental $t_{2g}/e_g$ peak ratio is smaller than unity, indicating other secondary excitation channels, such as electric dipole transitions (Refs. 24 and 25) or excitations to $p$ density of states (Ref. 10). Including these effects can modify the calculated intensity ratio but will not change the generic features.

[21] A. L. Ankudinov, B. Ravel, J. J. Rehr, and S. D. Conradson, Phys. Rev. B **58**, 7565 (1998); A. L. Ankudinov, C. E. Bouldin, J. J. Rehr, J. Sims, and H. Hung, *ibid.* **65**, 104107 (2002).

[22] Y. Ohta, T. Tohyama, and S. Maekawa, Phys. Rev. B **43**, 2968 (1991).

[23] J. Kuneš, A. V. Lukoyanov, V. I. Anisimov, R. T. Scalettar, and W. E. Pickett, Nature Mater. **7**, 198 (2008).

[24] M.-A. Arrio, S. Rossano, Ch. Brouder, L. Galoisy, and G. Calas, Europhys. Lett. **51**, 454 (2000).

[25] T. Yamamoto, X-Ray Spectrom. **37**, 572 (2008).